\documentclass[sigconf]{acmart}

\AtBeginDocument{%
  \providecommand\BibTeX{{%
    \normalfont B\kern-0.5em{\scshape i\kern-0.25em b}\kern-0.8em\TeX}}}

\copyrightyear{2024} 
\acmYear{2024} 
\setcopyright{rightsretained} 
\acmConference[CHI '24]{Proceedings of the CHI Conference on Human Factors in Computing Systems}{May 11--16, 2024}{Honolulu, HI, USA}
\acmBooktitle{Proceedings of the CHI Conference on Human Factors in Computing Systems (CHI '24), May 11--16, 2024, Honolulu, HI, USA}
\acmDOI{10.1145/3613904.3642839}
\acmISBN{979-8-4007-0330-0/24/05}

\usepackage{aksp}
\usepackage{array}
\usepackage{booktabs}
\usepackage{graphicx}
\usepackage{float}
\usepackage{soul,xcolor}

\newcommand{\sysname}{ShortScribe}
\setstcolor{red}

\begin{document}

\title{Making Short-Form Videos Accessible with \\ Hierarchical Video Summaries}


\author{Tess Van Daele}
\affiliation{
  \institution{Department of Computer Science \\ The University of Texas at Austin}
  \country{}}
\email{tessvandaele@utexas.edu}

\author{Akhil Iyer}
\affiliation{
  \institution{Department of Computer Science \\ The University of Texas at Austin}
  \country{}}
\email{akhil.iyer@utexas.edu}

\author{Yuning Zhang}
\affiliation{
  \institution{Department of Information Science \\ Cornell University}
  \country{}}
\email{yz3227@cornell.edu}

\author{Jalyn C Derry}
\affiliation{
  \institution{Department of Computer Science \\ The University of Texas at Austin}
  \country{}}
\email{jalyn.derry@utexas.edu}

\author{Mina Huh}
\affiliation{
  \institution{Department of Computer Science \\ The University of Texas at Austin}
  \country{}}
\email{minahuh@cs.utexas.edu}

\author{Amy Pavel}
\affiliation{
  \institution{Department of Computer Science \\ The University of Texas at Austin}
  \country{}}
\email{apavel@cs.utexas.edu}


\begin{abstract}
 Short videos on platforms such as TikTok, Instagram Reels, and YouTube Shorts (i.e. short-form videos) have become a primary source of information and entertainment. Many short-form videos are inaccessible to blind and low vision (BLV) viewers due to their rapid visual changes, on-screen text, and music or meme-audio overlays. In our formative study, 7 BLV viewers who regularly watched short-form videos reported frequently skipping such inaccessible content. We present ~\sysname{}, a system that provides hierarchical visual summaries of short-form videos at three levels of detail to support BLV viewers in selecting and understanding short-form videos. ~\revision{~\sysname{} allows BLV users to navigate between video descriptions based on their level of interest. To evaluate ~\sysname{}, we assessed description accuracy and conducted a user study with 10 BLV participants comparing ~\sysname{} to a baseline interface. When using ShortScribe, participants reported higher comprehension and provided more accurate summaries of video content.} 

\end{abstract}



\begin{teaserfigure}
    \centering
    \includegraphics[width=0.9\textwidth]{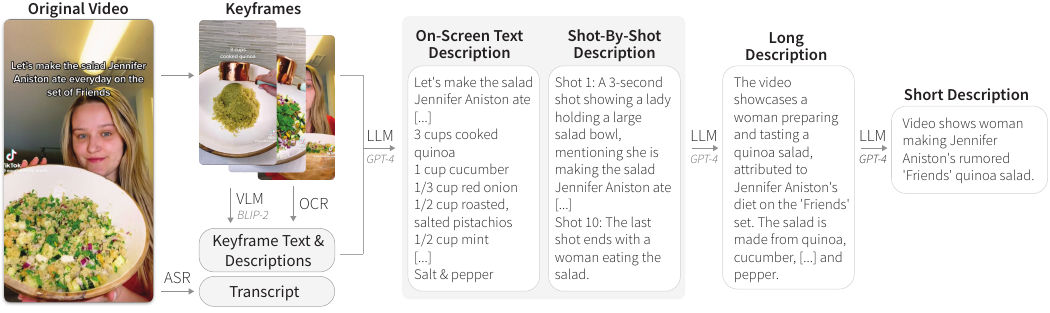}
    \caption{~\sysname{} makes short-form videos accessible with hierarchical video descriptions. ~\sysname{} extracts video data by identifying key frames then applying automatic speech recognition (ASR), automated description (BLIP-2), and optical character recognition (OCR). A large language model (GPT-4) then generates multiple descriptions. TikTok by @nourished.by.mads~\cite{nourishedbymads}.}
    \Description{The figure shows information from left to right. The most left is the original video represented as a single frame of a woman holding a bowl of salad with the text “Let's make the salad Jennifer Aniston ate every day on the set of Friends.” Two arrows point to the right from this picture. One points to the key frames which is a stack of single frames showing ingredients being added to a bowl. The other arrow points to the transcript of the video (transcribed with automatic speech recognition). 
    Below the keyframes are two arrows (one labeled with the vision to language model  BLIP-2) and one labelled OCR both pointing to "Keyframe Text \& Descriptions". 
    To the right of the key frames, key frame text \& description and transcript is an arrow (labelled LLM GPT-4) pointing right to two descriptions which includes on-screen text description (“Let's make the salad Jennifer Aniston ate … 3 cups cooked quinoa 1 cup cucumber 1/3 cup red onion 1/2 cup roasted, salted pistachios 1/2 cup mint … Salt & pepper”) and the shot-by-shot-description (“Shot 1: A 3-second shot showing a lady holding a large salad bowl, mentioning she is making the salad Jennifer Aniston ate…Shot 10: The last shot ends with the woman eating the salad”). 
    To the right of these two descriptions an arrow labelled LLM and GPT-4 points to a long description (“The video showcases a woman preparing and tasting a quinoa salad, attributed to Jennifer Aniston's diet on the 'Friends' set. The salad is made from quinoa, cucumber, …and pepper.”). To the right of the descriptions is an arrow also labelled LLM and GPT-4 pointing right to a short description (“Video shows woman making Jennifer Aniston's rumored 'Friends' quinoa salad.”).}
    \label{fig:teaser}
\end{teaserfigure}


\maketitle
\section{Introduction}
Today, more than 1 billion users actively watch short-form videos across platforms such as TikTok, Instagram Reels, and YouTube Shorts~\cite{tiktok-1billion}. 
Short-form videos usually range from 30 to 60 seconds in length and are presented in an algorithmically currated stream.
To capture viewer attention, creators make densely packed videos featuring rapid scene changes, on-screen text overlays, and fast-paced action. Creators also engage with trends by reusing popular music and meme audio, and by stitching or overlaying their responses to other videos. While short-form videos are now a dominant source of information, entertainment, and cultural references, they remain inaccessible to millions of blind and low vision (BLV) viewers.

Prior work has explored making videos accessible with manual~\cite{youdescribe}, semi-automated~\cite{pavel2020rescribe,branje2012livedescribe,yuksel2020human,liu2022crossa11y} and automated~\cite{wang2021toward,campos2020cinead} \textit{audio descriptions} --- or, narrations of visual content in the video. 
The duration and density of short-form videos make it challenging to fit audio descriptions within gaps in the audio.  
For videos without audio gaps, accessibility guidelines recommend adding extended descriptions that pause the video to narrate important visual content~\cite{wcag2}. While such extended descriptions can be useful for educational videos~\cite{peng2021slidecho}, they lengthen the content and abruptly interrupt the audio~\cite{pavel2020rescribe}. Gleason et al. explored how to make silent GIFs accessible by manually creating three types of descriptions (alt text, source audio, and source audio with audio descriptions), finding that audience members preferred alt text descriptions as they were efficient to read with their screen reader~\cite{gleason2020making}. Prior research and video accessibility guidelines have yet to explore how to make short-form videos accessible to blind and low vision viewers.

To understand the current practice of BLV viewers, we conducted formative interviews and a co-watching exercise with 7 BLV participants who regularly watched short-form videos. Participants reported frequently encountering videos with audio unrelated to the visual content of the video (\textit{e.g.}, trending songs, meme audio), making it difficult to determine what was happening on screen. 
Participants also reported that they did not know whether the video would be accessible ahead of time, thus having to watch the majority of a video just to find out it was inaccessible.
When encountering inaccessible videos, participants either skipped the video, asked a friend for more information, or posted the video on online communities to request descriptions if they were particularly interested in the content. Still, participants ended up skipping a majority of inaccessible videos and rarely followed up for more information, limiting their access to short-form content.

To make short-form videos accessible, we present ~\sysname{}, a system that provides BLV short-form video viewers with ~\revision{hierarchical video summaries, or video descriptions at multiple levels of detail that viewers can access depending on their interest.} 
To create these descriptions, our system first segments videos into shots (\textit{i.e.} camera or scene changes) and extracts visual information from each shot using vision language models (BLIP-2, OCR). Next, it uses a large language model (GPT-4) to summarize the extracted visual information, creating descriptions for each shot (shot-by-shot descriptions) and for the entire video (short description, long description, and on-screen text). Using the short description, participants can quickly determine if the content of a video is interesting to them, and then flexibly explore additional details through the long description, shot-by-shot descriptions, and on-screen text. 

We evaluated ~\sysname{} in a within-subjects study with 10 BLV participants who compared \sysname{} to a baseline interface created to simulate a typical short-form video platform. Participants demonstrated improved video comprehension and reported that they found unique uses for each of the descriptions we provided. Participants expressed unanimously that they would use \sysname{} in the future and that it improved their experience watching short-form videos. 

Our work contributes:
\begin{itemize}[noitemsep,topsep=0pt]
    \item A formative study revealing current practices and challenges of watching short-form videos for BLV users
    \item Design and development of \sysname{}, a system that provides BLV users with hierarchical visual descriptions 
    \item A user study demonstrating that \sysname{} improved the experience of watching short-form videos as well as selecting which videos participants wanted to watch. 
\end{itemize}

\section{Background \& Related Work}
\revision{Our work on making short-form videos accessible with hierarchical video descriptions relates to prior work in video accessibility, hierarchical summarization, and social media accessibility.}

\subsection{Video Accessibility}

\revision{While a long history of work has explored how to make videos accessible for BLV viewers, prior research has not yet explored how to make short-form videos accessible~\cite{nevsy2023audiovisualmedia}. We review primary approaches for making videos accessible: } 

\subsubsection{Audio Description} Videos are often inaccessible to blind viewers when the visual content cannot be understood from the audio alone~\cite{liu2021what}. Web Content Accessibility Guidelines (WCAG 2.0) recommend that to make videos accessible, authors can create a summary of the content or add audio descriptions that synchronously narrate the visual content while avoiding overlapping with important audio content~\cite{wcagad}. To support authors creating audio descriptions, prior work has proposed manual~\cite{youdescribe,killough2023exploring}, collaborative~\cite{natalie2021efficacy}, and (semi-)automated~\cite{natalie2023supporting,pavel2020rescribe,branje2012livedescribe,yuksel2020human,wang2021toward,gagnon2009towards,campos2020cinead,liu2022crossa11y} approaches to create descriptions for a range of videos including long-form traditional films and TV shows~\cite{gagnon2009towards,campos2020cinead}, user-generated videos~\cite{pavel2020rescribe,liu2022crossa11y,wang2021toward,natalie2021efficacy,natalie2023supporting,youdescribe}, livestreams~\cite{killough2023exploring, jun2021streamersblv}, and 360-degree videos~\cite{chang2022omniscribe, jiang2023360videoacc, fidyka2018audio360video}. 
Short-form videos often include continual audio such that clear gaps do not exist for adding audio descriptions.

When audio descriptions do not fit within audio gaps, the Web Accessibility Initiative suggests providing \textit{extended descriptions} that pause the underlying video so there is additional time to add descriptions~\cite{wcag2}. 
While extended descriptions may be well-suited for educational content, they introduce delays and confusing interruptions that are unsuitable for short-form videos. 
Prior work has explored prompting authors to add descriptions during recording~\cite{peng2021say} or providing users control over the playback of extended descriptions~\cite{pavel2020rescribe,peng2021slidecho,stangl2023dialogueagent}.
Rescribe let users control whether they receive inline, extended, or hybrid descriptions
ahead of time~\cite{pavel2020rescribe}, whereas other systems provided users on-demand access to extended descriptions~\cite{peng2021slidecho, stangl2023dialogueagent} or answers to their visual questions~\cite{stangl2023dialogueagent}.

All prior work on inline and extended audio descriptions focused on long-form videos and thus created \textit{time-aligned} descriptions, or descriptions that played back according to the time in the video (\textit{e.g.}, describing the setting at the beginning of a scene). To accommodate short-form videos, we explore text descriptions of the video that are not time-aligned to provide an overview of the video as a whole and detail on-demand.

\revision{\subsubsection{Text Descriptions and Summaries} Commercial and research tools have explored using text descriptions or summaries of a video support BLV and sighted audiences. Video platforms for long-form videos such as YouTube~\cite{youtube}, Vimeo~\cite{vimeo}, and Netflix~\cite{netflix}, display a text ``title'' (\textit{e.g.}, \textit{Backyard Squirrel Maze 1.0}) and text ``description'' (\textit{e.g.}, \textit{``Squirrels were stealing my bird seed so I solved the problem with mechanical engineering :)''}) for each video~\cite{backyardsquirrelmaze}. Creators use the text title, text description, and video thumbnail to preview the video content, and prior work noted that BLV audience members use text titles to select what video to watch~\cite{liu2021what}. Short-form videos lack such text titles that may provide a useful summary to BLV audiences. While short-form video platform (\textit{e.g.,} TikTok, Instagram Reels) creators can use the ``caption'' field to add text descriptions, this usually contains extra context about the video (\textit{e.g.}, the products displayed, the backstory behind the video, or an opinion on the topic) rather than describing the video content which might benefit BLV users but be repetitive to sighted audiences. Discussions about adding text descriptions for short-form videos exist~\cite{tiktok-descriptions}, but the practice is not widespread. We explore using automatically generated text descriptions to support BLV users gaining a quick overview of short-form videos.

Prior work has also explored text descriptions for GIFs (short, silent videos) and as a replacement for audio descriptions in long-form videos. Gleason et al. explored how to make GIFs accessible using three types of manually authored descriptions: alternative text description (most preferred by users), adding original source audio, and adding audio descriptions to original source audio~\cite{gleason2020making}. Similar to images, GIFs are silent and often convey a brief concept or scene (\textit{e.g.,} ``Oprah shrugs''). We explore short text descriptions for short-form videos, but as short-form videos convey more information than GIFs we use multi-modal summarization to take into account audio content and provide flexible access to additional descriptions. 
At the other end of the spectrum, for long-form videos, accessibility guides advocate for including \textit{descriptive transcripts} that describe visuals in the video and transcribe the audio into captions in the same document to support people who are Deaf-blind and others~\cite{descritivetranscripts, wcag2}. Prior work has explored authoring descriptions and captions~\cite{pavel2020rescribe,branje2012livedescribe,liu2022crossa11y}, but little prior work has explored creating descriptive transcripts. Similar to a descriptive transcript, AVScript~\cite{huh2023avscript} supports BLV video editors with an ``audio-visual script'' that included transcribed speech and descriptions of visual scenes. Building on such detailed descriptions, we explore enabling users to optionally access shot-by-shot descriptions that summarize the audio and visual details in each part of the video. 

\subsubsection{Beyond Descriptions} Videos can also be made more accessible using alternative modalities such as haptics or sound. 
Prior work has demonstrated the benefit of using haptics to conveying spatial information (\textit{e.g.}, location of actors on screen and facial expressions) in movies~\cite{viswanathan2010haptics} and 360-degree videos~\cite{jiang2023360videoacc}. Foley, or the reproduction of everyday sound effects added to videos to enhance audio quality, can artificially add rich information to the audio track of videos~\cite{foley}. Such approaches enable rich spatial and sonic experiences, but we explore text descriptions in this project as they are broadly usable with existing devices and screen readers.
}

\subsection{Heirarchical Summaries and Descriptions}
Prior work has used hierarchical summaries to support searching, browsing, and skimming of long text~\cite{zhang2017wikum,verroios2014context}, ~\revision{images~\cite{stangl2021onesizefitsall, huh2023genassist, lee2022imageexplorer},} audio recordings~\cite{li2021hierarchical,li2023improving}, and videos~\cite{pavel2015sceneskim,pavel2014video}.
Wikum and Context Trees both explore hierarchical summarization as an approach to crowdsource summarizations of text~\cite{verroios2014context} and help people skim discussions~\cite{zhang2017wikum}. 
Li et al. use bottom up hierarchical summarization to enable listeners to explore long-form dialog effectively, providing an overview and the ability to navigate to points of interest~\cite{li2021hierarchical,li2023improving}. SceneSkim provides video summaries at three levels of detail (plot summary, script, and captions), helping film professionals in searching and browsing for specific moments within a film~\cite{pavel2015sceneskim}. Video Digests provides chapter and section summaries for learners skimming and browsing lecture videos~\cite{pavel2014video}. While this prior work provides hierarchical summarization for text~\cite{zhang2017wikum,verroios2014context} and speech-based content~\cite{pavel2014video,pavel2015sceneskim,li2021hierarchical,li2023improving}, we consider providing hierarchical summarization of \textit{visual descriptions} of videos to facilitate gaining a high-level overview or low-level details about visual content. 

While ``a picture is worth 1000 words'' and vision language models can produce long descriptions~\cite{bard}, most BLV users do not want to read a 1,000 word description for every image. Accessibility research has started to explore using hierarchical image descriptions as an approach to support users in selectively navigating image details. Prior work uses vision language models to provide similarities and differences between images~\cite{huh2023genassist} or focus on specific parts of an image~\cite{lee2022imageexplorer}. As efficiency is important for short-form videos, we explore hierarchical summaries as an approach for users to flexibly explore video details.

\subsection{Social Media Accessibility}
\revision{The rise of social media (\textit{e.g.,} TikTok~\cite{tiktok}, Instagram~\cite{instagram}, YouTube~\cite{youtube}) has brought a rise in inaccessible images and videos. Prior work has investigated ways to support BLV social media users as authors~\cite{huh2023avscript, killough2023exploring, bennett2018how} and consumers~\cite{gleason2019making,gleason2020twitter,gleason2020making,killough2023exploring,wu2017automatic, liu2021what} of new types of images and videos used on social media. Researchers, practitioners, and advocates have iteratively released guidance on effective descriptions for images and video on social media~\cite{tiktok-descriptions, gleason2020making, killough2023exploring, gleason2019making}. We build on such prior work by exploring current practices of BLV short-form video audiences and exploring text descriptions for making such videos accessible.
Prior research has studied short-form videos to assess their impact on viewers~\cite{milton2023see, chiossi2023short, tan2022uncovering} and recent research explores accessibility issues of short-form videos to uncover opportunities to support neurodiverse viewers~\cite{simpson2023hey} and viewers with physical disabilities~\cite{duval2021chasing}. We explore the accessibility of short-form videos to uncover opportunities to support BLV viewers. 
}

\section{Formative Study}
Prior work has studied viewing practices, accessibility challenges, and solutions of BLV viewers for both long videos like TV shows~\cite{packer1997s}, livestreams~\cite{killough2023exploring}, and YouTube videos~\cite{liu2021what}, and short videos like silent GIFs~\cite{gleason2020making} commonly found on social media. We conducted a study to explore these topics for short-form videos, a new form of social media content not yet explored by prior work.

\subsection{Method}
The formative study included 7 BLV participants who accessed their mobile device using a screen reader and had experience watching short-form videos (Table~\ref{tab:pbackground-formative}). We recruited participants using social media and mailing lists. Participants were 20-57 years old and described their visual impairment as blind (5 participants) or some light perception (2 participants).

We asked participants a series of demographic and background questions including what types of short-form videos they watched, how they found those videos, the accessibility barriers that they encountered, and how they navigated those barriers. 
We then asked participants to watch 8 pre-selected short-form videos (Table~\ref{tab:pre-selected videos}). 
\revision{To ensure the 8 pre-selected videos covered a wide range of accessibility levels, we first created a 4-point scale to categorize the level of audio-visual match (as audio-visual mismatches indicate inaccessible video segments~\cite{liu2022crossa11y}): Unrelated Audio (\textit{e.g.}, a trending song), Somewhat Related Audio (\textit{e.g.}, a meme-style audio), Partially Informative Audio (\textit{e.g.}, recipe narration that covers some but not all details), and Mostly Informative Audio (\textit{e.g.}, a talking head). Videos were selected by scrolling through a newly created TikTok account, eliminating videos not in English or containing inappropriate content, until we found 4 videos per category. Finally, 2 videos were selected for each category such that the entire sample of videos covered a wide range of lengths and topics. }  
Participants watched the pre-selected videos in a random order, and then provided accessibility ratings from 1 to 7 (completely inaccessible to completely accessible), and explained what factors impacted their rating, similar to Liu et al.~\cite{liu2021what}. Finally, we invited participants to think aloud as they watched videos on their own short-form video feed with their preferred platform for 10 minutes (2 participants used TikTok, 1 Instagram Reel, 3 YouTube Shorts, and 1 Twitter). We conducted this 1.5 hour long study via 1:1 remote Zoom interviews and compensated participants \$37.50 via Amazon Gift Card. Participants were asked to screen-share during the interview. This study was approved by our institution’s IRB. 

To collect data, one of the authors took notes during the interviews. Another author re-watched the entire set of zoom recordings, adding to the notes and constructing an affinity diagram~\cite{Lucero2015ad}. The two authors later discussed themes that emerged in the affinity diagramming process to align on the results.\\

\begin{figure}[t]
    \centering
    \includegraphics[width=3.33in]{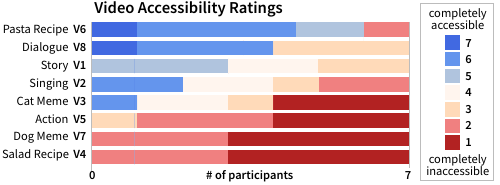}
    \caption{Participant ratings of video accessibility for pre-selected videos.}
    \label{fig:rating}
    \Description{The figure shows a horizontal bar graph of participant ratings of accessibility. The x-axis is the number of participants (1 to 7) and the y-axis shows videos which read (from top to bottom), “Pasta recipe V6, Dialogue V8, Story V1, Singing V1, Cat Meme V3, Action V5, Dog Meme V7, Salad Recipe V4.” The horizontal bars are filled with colors correlating to how the participants ranked the accessibility of the video, red being not accessible and blue being accessible. V7 and V4 scored the worst with all participants ranking them as either a 1 or 2. V6 and V8 scored the best.}
\end{figure}

\subsection{Findings}
\subsubsection{Current Practice} Participants watched short-form videos daily or weekly for entertainment (P1, P2, P4), staying up to date on popular trends (P4, P6), following creators (P3, P7), ~\revision{engaging with content shared by friends (P1, P2, P3, P4, P5, P7)}, and seeking information about specific interests (P1, P4, P5) like music, politics or education. P6 highlighted the importance of short-form videos being brief in length such that you quickly capture key information: \textit{``when I first started, I watched longer videos. But I find now I move to shorter videos, videos that get to the point quickly.''} All participants faced accessibility barriers when accessing short-form videos. When encountering these barriers, most participants skip the video (P1, P2, P3, P6, P7), but others seek help from close friends (P2, P4, P5) or social media groups that provide video descriptions for the BLV community (P2, P3).
Participants also mentioned using the author-written video caption (P1, P6) or searching online (P6) to gain more information. P4, P5, and P7 expressed frustration when they failed to find more information to better understand the video, feeling that they might miss crucial content.

\subsubsection{Short Form Video Accessibility}
Overall, participants rated the accessibility of the pre-selected videos as a 3.21 ($\sigma=1.94$) on a scale of 1 (completely inaccessible) to 7 (completely accessible) (Figure~\ref{fig:rating}).
Similar to Liu et al.~\cite{liu2021what}, participants noted that videos with more speech such as stand-up comedy or podcast excerpts (P2, P4, P7), educational videos (P1), and singing or music related videos (Figure~\ref{fig:rating}, V2) were more accessible than those with less speech. Short-form videos also presented unique accessibility challenges:

\noindent \textbf{Repurposed Audio: } Participants reported that short-form videos that reused audio from other sources were challenging to understand. In some cases, the audio was somewhat related to the visual content, but still did not contain enough information for participants to understand the video. V3 and V7 both used audio memes. In V3, the video depicts a cat yawning, the audio features a woman saying \textit{``Mm mm mm (disapproval), today drained me''}, the text on screen says \textit{``My cat sleeping for 18 hours, stomping on my head at 4am..., and beating up his brother for no reason''} and the caption states \textit{``He works so hard''}. In V7, the video features a dog cuddling chipmunk stuffed animals and the audio is a woman singing \textit{``If you're not real, how come I feel this way? Little babies.''} With access to only the audio and author written captions, participants commonly misinterpreted the subject of these two videos. For instance, P4 interpreted V3 as a fatigued woman shaking her head, and V7 as a person with a baby. For a video depicting a recipe montage set to entirely unrelated music (V4), participants recognized the video as being inaccessible (Figure~\ref{fig:rating}), and were not misled by the audio. As P1 shared: \textit{``when it’s just like music and images I just skip past that because that’s not accessible''}. Due to the presence of repurposed audio on the platform, participants occasionally guessed audio was repurposed when it was not. For example, for a video with a woman singing a song to the camera (V2), P6 asked if the video was of a person lipsyncing to a song.

\noindent \textbf{Micro Videos}: Participants indicated that extremely short videos (5 seconds or less) often had uninformative or ambiguous audio that interfered with screen reader audio, making these videos inaccessible. Participants thus rated videos 5 seconds or less (V3, V5, V7) with accessibility scores of 3 or less (Figure~\ref{fig:rating}).
~\revision{
While V3 and V7 featured meme-style repurposed audio, V5 included original audio of a man saying ``no!'' then a splash. Participants were unsure what happened in the video (a dog jumped into a pool onto a man). 
All videos play in a continuous loop (\textit{e.g.}, V5 had a ``no'' then splash loop) and participants reported that the looping audio was overwhelming and overlapped with the audio of their screen reader such that they needed to pause the video to look for more information, disrupting their browsing.} 

\noindent \textbf{Reaction Videos}: P6 reported that reaction videos in which one creator stitched the video of another creator to react to it, were difficult to understand from audio alone. P6 shared that it would be challenging to add descriptions for such videos: \textit{``Multi-layers of audio description would be needed, you know. What's the original person doing? And how does that match up with you?''} (P6). 

\noindent \textbf{Complex Actions}: P7 reported that short videos with excessive animations and movements were not accessible, as these videos are unlikely to be adequately described in limited time. \revision{For example, P5 said that \textit{``Dance is always inaccessible''}, as the complex, fast-paced movements and changing facial expressions in dance make it difficult to fully describe.
P6 explained that even a narration referencing primarily visual content, such as \textit{``birds flying half a mile away,''} signifies inaccessibility. While the narration references the visual focus for sighted viewers, it does not provide sufficient descriptions of visual details for BLV viewers to also appreciate the visuals (\textit{e.g.}, movements of the birds, appearance of half a mile).} 

\subsubsection{Platform Accessibility} Participants rated platform accessibility as 3.36 ($\sigma=1.49$) on a scale of 1 to 7 (completely inaccessible or accessible, respectively) and highlighted key accessibility barriers: 

\noindent \textbf{Video Controls:} All participants reported that play/pause video controls were challenging to use on TikTok as they required tapping the screen, a gesture that was not supported when using VoiceOver. Several participants also noted difficulties with other playback controls, such as skipping, fast-forwarding, and rewinding. 
\revision{Most short-form video platforms loop video playback by default, posing accessibility challenges. When a loop occurred, it was challenging for participants to tell from the audio alone if the application was still playing the video, had switched to the next video, or if the video had looped.} 
Participants expressed a preference for proactive video control, rather than automatic video playback, and noted that video controls on YouTube Shorts were the most accessible.

\noindent \textbf{Button labels}: All participants reported that clutter and a lack of clarity in button labels led to the platform being inaccessible. As P6 described: \textit{``
The comment buttons and the share buttons, I don't know which video even they are connected to.
I may find a play button, but it's not necessarily the one for the video that I'm trying [to play].''}

\noindent \textbf{Platform Updates}: Five participants noted that updates in the platform layout, particularly changes in the button positions, 
incurred a steep learning curve, leading to moments of inaccessibility.

\subsubsection{Participant Suggested Accessibility Improvements}

Participants suggested adding descriptions for short videos (6 participants) and providing access to the text on-screen (5 participants) to improve understanding of the short-form video content. Three participants suggested making it easier to access the author-written video caption and user comments, as these helped participants decide whether or not to watch a video. Two participants suggested that developing additional VoiceOver gestures for fast navigation.

\subsection{Design Implications}
Our formative study revealed design opportunities for making short-form video viewing and browsing accessible:

\begin{itemize}
    \item[\textbf{D1.}] Provide efficient access to on-screen visuals and text.
    \item[\textbf{D2.}] Support users in recognizing audio and visual mismatches.
    \item[\textbf{D3.}] Enable screen reader control of video playback and browsing.
    \item[\textbf{D4.}] Support users in deciding on a video to watch.
    \item[\textbf{D5.}] Maintain fast-paced video viewing and browsing.
    \item[\textbf{D6.}] Provide access to complex visual content (\textit{e.g.}, dance). 
\end{itemize}

We support BLV short-form video viewers with short, long and shot-by-shot descriptions that we crafted according to these design goals. First, to enable screen reader users to efficiently decide whether or not to watch a video (\textbf{D4}), we provided \textbf{\textit{short descriptions}} that share a brief summary of video content (\textbf{D1}). Such short descriptions provide a similar function to long-form video titles, as prior work indicated that video titles support BLV users in selecting videos to watch~\cite{liu2021what}. 
While short descriptions can support users quickly selecting videos (\textbf{D4}) or noticing audio-visual mismatches (\textbf{D2}), short descriptions optimize for efficiency (\textbf{D5}) over completeness and thus may leave out important information. We enable BLV users to flexibly gain detailed access to audio-visual content and on-screen text (\textbf{D1}) with \textbf{\textit{long descriptions}}, \textbf{\textit{shot-by-shot descriptions}}, and a transcript of \textbf{on-screen text}. 
\revision{We prioritize displaying the short description first in the interface, and enable flexible description access to longer descriptions, such that users can decide whether or not they want to spend additional time to read longer descriptions (\textbf{D5}). Reading longer descriptions can support users in clarifying misunderstandings from mismatched audio, or even recognizing more subtle audio-visual mismatches such an ingredient that a recipe narration left out (\textbf{D2}). We create our interface with explicit video playback control such that the interface is accessible and efficient to use for screen reader users (\textbf{D3}, \textbf{D5}).}
Our work does not address \textbf{D6}, but it offers a rich opportunity for future work.

\section{System}

We present \sysname{}, a system that makes short-form videos accessible to BLV users by providing hierarchical descriptions of short-form videos.
\begin{figure}
    \centering
    \includegraphics[width=3.33in]{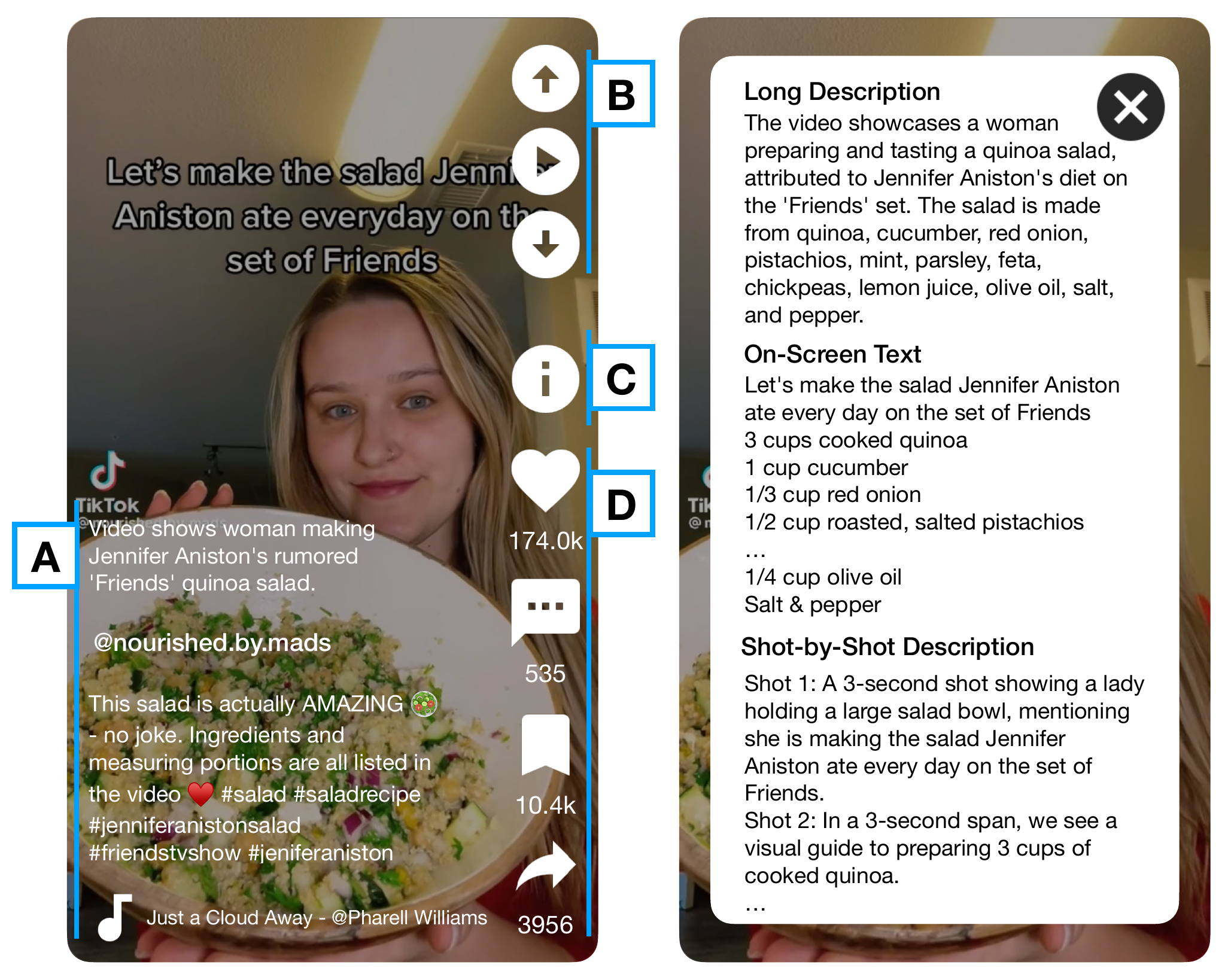}
    \caption{The ~\sysname{} interface consists of (a) front screen video information including the short description, username, caption, and audio title, (b) video controls, (c) a button to open the description pane which includes the long description, on-screen text, and shot-by-shot descriptions, and (d) video statistics. Video Credit: TikTok used with permission from @nourished.by.mads~\cite{nourishedbymads}.}
    \label{fig:interface}
    \Description{The figure shows two screens of a mobile interface, one on the left and one on the right. Left: A still frame of a woman holding a bowl of salad. Within the video, there is text that reads “Let's make the salad Jennifer Aniston ate every day on the set of Friends.” Elements on top of the video are grouped into four groups A, B, C, and D. In group A, in the bottom left, video text information is displayed (marked with an A) including a short description (“Video shows woman making Jennifer Aniston's rumored 'Friends' quinoa salad.”), username (“@nourished.by.mads”), video caption (“This salad is actually AMAZING - no joke. Ingredients and measuring portions are all listed in the video #salad #saladrecipe #jenniferanistonsalad #friendstvshow #jeniferaniston”), and audio title (“Just a Cloud Away - @Pharell Williams”). The screen also includes button icons aligned vertically along the right of the screen. Group B is the top three buttons which includes previous (upward arrow), play/pause (play icon), next (downward arrow). Group C includes the video description button (information icon). Group D includes like (heart), comment (speech bubble), bookmark (save icon), and share (share icon). Right: A popup is shown that presents three descriptions and a close button in the top right corner. The text includes a long description (“The video showcases a woman preparing and tasting a quinoa salad, attributed to Jennifer Aniston's diet on the 'Friends' set. The salad is made from quinoa, cucumber, red onion, pistachios, mint, parsley, feta, chickpeas, lemon juice, olive oil, salt, and pepper.”), on-screen text (“Let's make the salad Jennifer Aniston ate every day on the set of Friends 3 cups cooked quinoa 1 cup cucumber 1/3 cup red onion 1/2 cup roasted, salted pistachios … 1/4 cup olive oil Salt & pepper”), and shot-by-shot-descriptions (“Shot 1: A 3-second shot showing a lady holding a large salad bowl, mentioning she is making the salad Jennifer Aniston ate every day on the set of Friends.Shot 2: In a 3-second span, we see a visual guide to preparing 3 cups of cooked quinoa. …”).}
\end{figure}

~\revision{
\subsection{Viewing Videos with \sysname{}}
Mari is blind and recently started cooking more often. She likes to find ideas for new recipes by watching short-form videos on TikTok. While scrolling, Mari comes across a video set to a popular song (Figure~\ref{fig:interface}, left). To find out what the video is about, Mari uses her screen reader to read the short description that states the video features a woman making a quinoa salad (Figure~\ref{fig:interface}A). As Mari is looking for easy and filling lunch options, she checks out the long description to read a more detailed description that shares a high-level overview of the ingredients (\textit{e.g.}, ``pistachios, mint, chickpeas''). She decides she wants to make the salad later, so she reads the on-screen text, which specifies detailed ingredient amounts (\textit{e.g.}, ``3 cups cooked quinoa''). She copies the ingredients into her shopping list, and reads the shot-by-shot description to find how the creator prepares the salad (adds ingredients one by one to a large bowl). Mari continues to scroll to another video and reads the short description to notice this video is about skincare products (a topic she is not interested in), so she immediately scrolls to the next video in her feed. 
}

\subsection{Interface}
\revision{We created a mobile interface to provide access to \sysname{}'s hierarchical video summaries (Figure~\ref{fig:interface}). \sysname{} features a \textit{video pane} that mimics existing short-form video platforms. We designed the video pane to be screen reader accessible by modeling the design off of YouTube Shorts as formative study participants highlighted it as the most accessible platform, and further modifying the design according to suggestions in the formative study (\textit{e.g.}, do not loop the video). We added \sysname{}'s short descriptions to this video pane. \sysname{}'s \textit{description pane} is unique to our interface and lets users flexibly gain access to additional information via long descriptions, shot-by-shot descriptions, and on-screen text transcripts. Our interface serves to demonstrate how \sysname{}'s hierarchical video summaries may be integrated into existing short-form video platforms.}

\subsubsection{Video Pane} The video pane features the current video, video controls, and additional information about the video (\textit{e.g.} short description, username, caption, number of likes).
To switch between videos, users navigate to the previous or next buttons with their screen reader and double-tap the buttons. To play and pause a video, users double-tap the play button. The video pane displays information about the video in the following read order: short description, username, caption, source audio title, description pane button, number of likes, number of comments, number of bookmarks, and number of shares. The video pane directly displays the short description. The long descriptions, on-screen text, and shot-by-shot descriptions are included in the description pane.

\subsubsection{Description Pane} The description pane enables users to flexibly gain access to additional in-depth information about the video. In the description pane, the user can access the video's long description, on-screen text, and shot-by-shot descriptions. The long video description provides a paragraph summarizing the  video based on data such as on-screen text, audio transcription, and vision-to-language model-generated image captions. Users can also directly access on-screen text extracted from the video which is particularly useful for videos where creators included this type of text(\textit{e.g.}, recipes that list the ingredients on-screen). Finally, users can access the shot-by-shot descriptions which include descriptions of each shot in the video and provides more granular information on the video. To go back to the video pane after reading this information, users can double tap the close button.

\subsubsection{Implementation}

The interface (Figure ~\ref{fig:interface}) for \sysname{} was implemented using React.js (front-end) and real-time Firebase database (back-end) that logged user interactions for button clicks. All descriptions created from the pipeline were pre-loaded into JSON files and integrated into the React.js code base to filter out inappropriate words in the descriptions. We tested to ensure the interface was compatible with popular screen readers such as Apple VoiceOver and NVDA.

\begin{figure}
    \centering
    \includegraphics[width=3.33in]{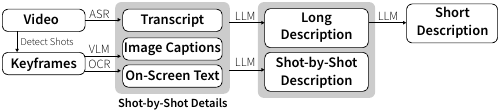}
    \caption{\sysname{} takes a video as input, transcribes the audio using automatic speech recognition (ASR), segments the video into shots, and selects the middle frame of each shot as a keyframe. It then processes the transcript, generated image captions (BLIP-2), and on-screen text (OCR) to produce video data for each keyframe. We use a large language model (GPT-4) to summarize this data into a short, long, and shot-by-shot description.}
    \label{fig:system-diagram}
    \Description{The figure shows a diagram of how the system works. From left to right, the diagram shows 4 groupings or columns, all connected by arrows pointing right. The first group includes Video and Keyframes with an arrow pointing downward from Video to Keyframes. From Video, there is an arrow that says ASR pointing right to Transcript which is in the second group. From Keyframes, there is an arrow that says VLM pointing right to Image Captions and another arrow that says OCR pointing right to on-screen text. Transcript, Image Captions, and on-screen text make up group 2 and are labeled at shot-by-shot details. From group 2, there are two arrows both saying LLM pointing right to long description and shot-by-short descriptions. From group 3, there is one arrow saying LLM pointing right to short description.}
\end{figure}

\subsection{Pipeline}
\subsubsection{Extracting audio and visual information.} 

Given a video, we first transcribe the audio in the video using Google Cloud's Speech Transcription API~/cite{googlespeech}. We then segment the video into shots, or continuous visual content segments using FFMPEG's SceneDetect~\cite{ffmpeg} to recognize sudden luminance changes. For each shot, we selected the middle frame as the representative keyframe. For example, if a shot is 30 frames long, we select the 15th frame as the keyframe. For each keyframe, we detected any on-screen text embedded in the video using Optical Character Recognition (OCR) from Google's Video Intelligence API~\cite{googlevideointelligence}. We filtered out any extracted text with confidence less than 0.95 to capture added on-screen text but avoid extraneous text (\textit{e.g.}, signs in the background of a video). We also filtered out content captured from the TikTok watermark (\textit{e..g.}, usernames and the word \textit{``TikTok''}). We then used the BLIP-2 XXL image captioning model~\cite{blip2} to generate 5 candidate descriptions for each keyframe. The model generated the candidate descriptions with nucleus sampling, a required length of 5 to 20 words, a top-p value of 0.9, and a temperature of 1. 
To select a final description from the 5 generated candidate descriptions, we computed the CLIP score~\cite{clip} between each candidate description and the shot keyframe, selecting the visual description with the highest score. 

\subsubsection{Generating Descriptions}
For each shot, we gather the visual description of the shot's keyframe, extracted on-screen text, and the corresponding audio transcript. To create the \textbf{shot-by-shot descriptions}, we prompt GPT-4~\cite{gpt4} to write a summary of each shot. We provide the audio and visual information for all of the shots at once to GPT-4 in order to provide context for each individual shot summary (full prompt in Appendix~\ref{shot_by_shot_prompt}). To generate the \textbf{long descriptions}, we similarly prompt GPT-4 to summarize the audio and visual information in all of the shots into a single summary paragraph (full prompt in Appendix~\ref{long_desc_prompt}). If the long description exceeds 50 words, we condense the long description into a 50-word description and use the 50-word description as the long description for the video (full prompt in Appendix~\ref{50wordprompt}). Finally, we create the \textbf{short descriptions} by prompting GPT-4 to condense the long description with the prompt: \textit{``Condense the summary below such that the response adheres to a 10 word limit.''}  

\section{Pipeline Evaluation}

\begin{figure}
    \centering
    \includegraphics[width=3.33in]{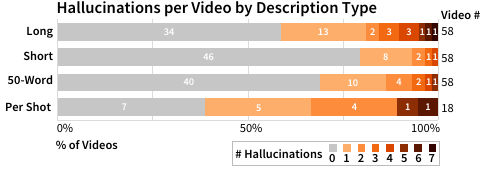}
    \caption{We analyzed hallucinations in descriptions for 58 videos (long, short, 50-word descriptions) and for a subsample of 18 videos (per shot descriptions). Descriptions for each video contained 0-7 hallucinations. Short descriptions had the lowest percentage of videos with hallucinations, while shot-by-shot descriptions had the highest percentage of videos with hallucinations.}
    \label{fig:hallucinations}
    \Description{A horizontal bar graph of hallucination made by the pipeline. The x-axis is the percentage of videos. The y-axis on the left is description type, which lists (from top to bottom) long, short, 50-word, and per shot. The y-axis on the right is how many videos were analyzed for each description type, which lists (from top to bottom) 58, 58, 58, 18. Short descriptions had the highest percentage of videos with no hallucinations, having 40 out of 58 with no hallucination. The per-shot descriptions had the lowest percentage of videos with no hallucinations, having 7 out of 18 with no hallucination.}
\end{figure}

\begin{table}[]
\centering
\resizebox{\columnwidth}{!}{%
\begin{tabular}{lllllllll}
\toprule
& \multicolumn{3}{l}{Hallucinations} & \multicolumn{2}{l}{Coverage} & \multicolumn{2}{l}{Words} \\ 
Description Type & \multicolumn{1}{l}{$\mu$} & \multicolumn{1}{l}{$\sigma$} & \multicolumn{1}{l}{\#} & \multicolumn{1}{l}{$\mu$} & \multicolumn{1}{l}{$\sigma$} &  \multicolumn{1}{l}{$\mu$} & \multicolumn{1}{l}{$\sigma$}\\ \midrule
Short Description &  0.33 & 0.78 & 19 & 75\% & 22\% & 10 & 1\\ 
50-Word Description &  0.57 & 1.08 & 33 & 90\% & 20\% & 43 & 5\\ 
Long Description & 0.97 & 1.63 & 56 & 100\% & 0\% & 136 & 37\\ 
Shot-by-Shot Description &  1.44 & 2.01 & 26 & 100\% & 0\% & 190 & 186 \\ \bottomrule
\end{tabular}%
}

\caption{The mean ($\mu$) and standard deviation ($\sigma$) of hallucinations and percent summary coverage for each description type. We also report the total number of hallucinations and summary points covered per description type. Hallucinations consider 58 videos (short, 50-word, and long descriptions) or a subset of 18 videos (shot-by-shot). Coverage considers the 8 pre-selected videos used in user study. Words considers all 58 videos.}
\label{tab:hallucination-stats}
\end{table}

We measured the coverage and accuracy of the short descriptions, long descriptions, shot-by-shot descriptions, and BLIP-2 generated image descriptions from \sysname{}.

\subsection{Dataset} 
We first selected a set of 58 short-form videos. We selected the videos by creating a new TikTok account to remove any viewing history, then scrolling through the suggested videos to select videos spanning a variety of general audience genres: pets, memes, dance, music, and recipes. The selected short-form videos were 5 to 90 seconds in length and we collected 5 to 15 videos per genre. We selected a subset of 8 of these videos (Table~\ref{tab:user-study-videos}) to represent a variety of accessibility levels (\textit{e.g.}, somewhat related audio, fully narrated) and wrote detailed summaries of the visual content for each video (i.e. human summaries). We used these summaries to evaluate coverage in the pipeline evaluation, and participant comprehension in the user study of the system. We ran our pipeline on all 58 videos, and included input data and pipeline results for all 58 videos in the Supplemental Material.

\revision{
\subsection{Analysis}
We evaluated the accuracy and coverage of \sysname{} descriptions. For accuracy, we assessed short, 50-word, and long descriptions for all 58 videos, and shot-by-shot descriptions for a subset of 18 videos (8 used for the user study, plus 10 randomly sampled). For coverage, we assessed short, 50-word, long, and shot-by-shot descriptions for the 8 videos for which we had human-written summaries.

To evaluate the accuracy of descriptions, we examined the descriptions and videos to tally the number of inaccurate statements in each video (\textit{e.g.}, the description contained ``guinea pig'' when no evidence of a guinea pig was present in the visuals). We consider statements in the descriptions incorrect if they could not be justified by any visual content, audio content, or on-screen text content. For shot-by-shot descriptions, we consider statements within each shot incorrect if they could not be justified by any content within their corresponding timestamps. To evaluate coverage of the descriptions, we compared the descriptions to the human-written summaries by tallying the number of important details mentioned by both the description and summary. 
One researcher labeled all data reported in the pipeline evaluation. To verify inter-rate reliability of our accuracy and coverage codes, a second researcher also labelled the data and we computed weighted Cohen's kappa for each description type. Agreement was ``moderate'' to ``almost perfect'' ($\kappa=0.50-1.0$) for accuracy and coverage across all description types except shot-by-shot description accuracy that had ``fair'' agreement ($\kappa=0.24$)~\cite{mchugh2012interrater}. In examining differences for shot-by-shot descriptions, researcher counts of inaccurate statements differed by 1.16 on average ($\sigma$=1.38). As shot-by-shot descriptions were much longer than other descriptions (averaging 190 words, Table~\ref{tab:hallucination-stats}) disagreements primarily occurred due to one researcher noticing an error the other missed and vice versa, rather than disagreement in the established code. }

\subsection{Accuracy}

Overall, a majority of short, 50-word and long descriptions did not contain incorrect statements (Figure ~\ref{fig:hallucinations}).

\begin{figure}
    \centering
    \includegraphics[width=3.33in]{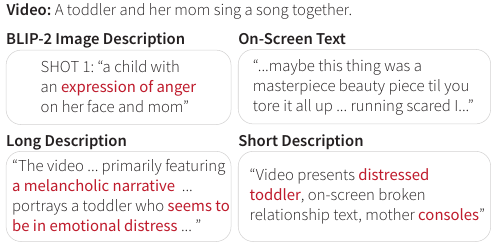}
    \caption{An analysis of the errors in one of the 2 of 58 videos that had more than three errors in the short description. The video depicts a lighthearted singalong. BLIP-2 mistakenly recognizes a toddler concentrating on singing as angry, and the on-screen text shows a quiz with the lyrics to to a sad song (\textit{All Too Well} by Taylor Swift). The long description and then short description incorrectly infer that the video is sad.}
    \label{fig:error-analysis}
    \Description{Video descriptions annotated for errors. At the top, the text reads “Video: A toddler and her mom sing a song together.” Next, each of the descriptions is shown in a grid (2 rows, 2 columns) with the following annotations: Blip-2 Image Descriptions. “Shot 1: a child with an [begin red text] expression of anger [end red text] on her face and mom” On-Screen text. “...maybe this thing was a masterpiece beauty piece til you tore it all up…running scared I…” Long description “The video …primarily featuring [begin red text] a melancholic narrative [end red text] portrays a toddler who [begin red text] seems to be in emotional distress [end red text]...” Short description “Video presents [begin red text] distressed toddler [end red text], on-screen broken relationship text, mother [begin red text] consoles [end red text].”}
\end{figure}

For long descriptions, 34 of the 58 videos contained no incorrect statements. Additionally, 13 of 58 long descriptions contained only one error, and 11 of 58 long descriptions contained more than one error (Figure ~\ref{fig:hallucinations}).
We investigated the long descriptions with more than three errors and found that the videos corresponding to these descriptions have no clear, informative audio or on-screen text content. For example, a video showcasing an unconventional apple peeler gadget without any useful description or on-screen text captions. The BLIP-2 model could not correctly identify what the gadget was doing based on the key frames. For instance, in one of the generated descriptions, BLIP-2 interpreted the gadget as a \textit{``mint toothbrush holder,''} resulting in the long description mentioning an \textit{``electric toothbrush device.''}

For short descriptions, there were no incorrect statements for 44 of the 58 videos. Additionally, only 3 videos had more than one incorrect statement (Table~\ref{tab:hallucination-stats}). The video with the greatest number of incorrect statements was a parent and their child performing a sing-along to a popular Taylor Swift song. BLIP-2 interpreted the child’s emotions as distressed due to their facial expressions. Additionally, the audio transcript captures the lyrics of the song about a broken romantic relationship. As a result, the long description and short description completely misunderstood the tone of the video as somber rather than humorous (Figure ~\ref{fig:error-analysis}). 

For the condensed 50-word descriptions, there were no incorrect statements for 40 of the 58 videos (Table~\ref{tab:hallucination-stats}). We found that the video with the greatest number of incorrect statements for this description type was also the child sing-along video due to the inaccurate shot descriptions and misleading audio. We also found that the shot-by-shot descriptions struggle to accurately describe \revision{}{videos with intricate visual details.} For example, in a soup recipe video, the image captions incorrectly identify several ingredients, resulting in 6 incorrect statements throughout the shot-by-shot descriptions.

\subsection{Coverage}
Overall, long descriptions and shot-by-shot descriptions captured all of the important details that we identified for the eight videos (Table~\ref{tab:hallucination-stats}). The short and 50-word descriptions capture 75\% and 90\% of the important details respectively. 
\sysname{} descriptions are comparable to human video descriptions in terms of coverage, but \sysname{}'s are generally more verbose than human descriptions. For example, one video is a person in a car zooming the camera in on a street sign reading \textit{``Drury Ln.``} and yelling out \textit{``The Muffin Man``} as a reference to the children's song. \sysname{} generated a long description that included the following segment: \textit{``The video begins with a car driving through a scenic route surrounded by trees and hills, and the on-screen text introduces the location as DRURY LN. The audio features a voice expressing surprise amongst the ambient noise of the driving car. The second shot, lasting four seconds, focuses on a road sign clearly showing the name 'DRURY LN', reiterating the location. However, there's no accompanying text or audio in this particular shot.''} The description mentions the street name twice and includes an unnecessary final sentence. 
\begin{figure*}
    \centering
    \includegraphics[width=\textwidth]{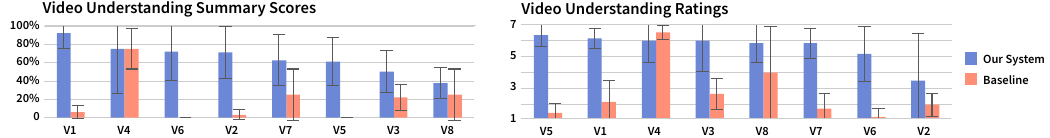}
    \caption{Video comprehension for videos V1-V8 using our system (left, blue) and a baseline interface (right, orange) measured by scoring participant written video summaries (Video Summary Scores) and participant's ratings of their video understanding (Video Understanding Ratings). Ratings of the video understanding ranged from 1, did not understand, to 7, completely understood. Error bars depict the 95\% confidence interval.}
    \label{fig:video-understanding}
    \Description{The figure shows two figures. 
    Left: A vertical bar graph titled “Video Understanding Summary Scores.” The x-axis is video title, listing (from left to right) V1, V4, V6, V2, V7, V5, V3, V8. The y-axis is the average score participants got on their summaries of the videos, listing (from top to bottom) 100\%, 80\%, 60\%, 40\%, 20\%, 0. Each video has two bars, the one on the left being blue (representing participants who used our system) and the one on the right being red (representing participants who used the baseline). V1 has the most significant difference between the system which scored 92\% and the baseline which scored 6\%. V4 had the least difference where the system and baseline both scored 75
    Right: A vertical bar graph titled “Video Understanding Ratings.” The x-axis is video title, listing (from left to right) V5, V1, V4, V3, V8, V7, V6, V2. The y-axis is the average accessibility rating participants gave each on a scale from 1 (not accessible) to 7 (accessible), listing (from top to bottom) 7, 5, 3, 1. Each video has two bars, the one on the left being blue (representing participants who used our system) and the one on the right being red (representing participants who used the baseline). V5 has the most significant difference between the system which scored 6.3 and the baseline which scored 1.5. V4 had the least difference between the system which scored 6 and the baseline which scored 6.5.}
\end{figure*}

\begin{figure}
    \centering
    \includegraphics[width=3.33in]{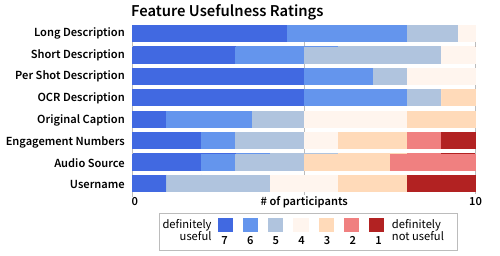}
    \caption{Participants rated the usefulness of each feature for understanding the video. The description features (first four features) are provided by \sysname{} only, and the remaining features (last four features)  were originally available on the short-form video platform. Engagement numbers refers to the number of likes, comments, and bookmarks.}
    \label{fig:feature-ratings}
    \Description{A horizontal bar graph titled Feature Usefulness Ratings. The x-axis is the number of participants (1 to 10) and the y-axis shows features which read (from top to bottom), “Long description, short description, per shot description, ocr description, original caption, engagement numbers, audio source, username.” The horizontal bars are filled with colors correlating to how the participants ranked the usefulness of each feature, red being definitely not useful and blue being definitely useful. The first four descriptions had overall high ratings while the last four had much lower ratings. }
\end{figure}

\section{User Evaluation}
We conducted a user study with 10 BLV participants to
examine how our descriptions impact video comprehension, selection, and preference compared to a baseline interface.

\subsection{Method}
In a within-subjects study, participants used both ~\sysname{} and a baseline interface to watch short-form videos. The study was 1.5 hours long, conducted in a 1:1 session via Zoom, and approved by our institution’s IRB. We compensated participants $\$37.50$ USD. \\

\noindent \textbf{Participants.} We recruited 10 BLV participants (8 female and 2 male) who use a screen reader to access their mobile device by using mailing lists and Facebook groups (Table~\ref{tab:pbackground-userstudy}). Participants ranged from 27 to 68 in age and described their visual impairment as blind (6 participants) or some light perception (4 participants). 9 out of 10 participants had experience watching short-form videos in the past although this was not a requirement for the study. P2, P3, P7, and P9 participated in the formative study. \\ 

\noindent \textbf{Materials.} The study consisted of two tasks: a video comprehension task and a video selection task.
For the comprehension task, we selected 8 videos from our 58-video dataset (Section 5.1) to represent a range of accessibility levels and split the videos into two groups (VG1 and VG2) such that each group had videos with similar levels of accessibility (Table~\ref{tab:user-study-videos}).
Each group had a video with reused meme audio, a video with audio original to the creator, a recipe video with limited auditory description, and a talking head of a person listing out items within a theme.
For the video selection task, we used the remaining 50 videos in the dataset and randomly divided them into two groups (VG3, VG4).
For each task, participants viewed one video group with the ~\sysname{} interface and one video group with the baseline interface. We randomized and counterbalanced the videos and interface pairs within each task. We also randomized the order in which they viewed the interfaces within each task. For both of the interfaces, we included information originally available on YouTube Shorts (\textit{e.g.}, author-written captions, author username, audio title, \# likes, comments, and bookmarks) along with the videos. For the ~\sysname{} interface, we added the descriptions generated in the pipeline (short, long, and shot-by-shot) as well as on-screen text. \\

\noindent \textbf{Procedure.} 
We asked participants demographic and background questions about their experience with short-form videos (if any), and then provided a 10-minute tutorial on how to use both the baseline and our system’s interface.
The rest of the study was split into two tasks:

\textit{Video Comprehension Task.} Participants watched 4 pre-selected short-form videos with the baseline interface and 4 videos with the system interface. 
For each video, we allowed participants to investigate as much information as they wanted and watch the video as many times as they wanted. 
We then asked participants to provide a short summary of the video and share any questions they had about the video. We also asked participants to rank their understanding of the video on a scale from 1 (did not understand) to 7 (completely understood).

\textit{Video Selection Task.} Participants were given 25 pre-selected short-form videos with the baseline interface and 25 videos with the system interface. Participants were asked to freely engage with the sample feed, similar to how they would typically engage with short-form video platforms. We asked them to think-aloud as they scrolled through the feed and optionally watched the videos. This task lasted until participants reached the end of the 25 videos or 15 minutes passed. 

At the end of the study, we conducted a final interview to understand participants’ experience with our system and asked participants to rate the usefulness of system features. \\

\noindent \textbf{Analysis.} We recorded and transcribed the interviews and recorded all interactions with both interfaces. \revision{To examine participant summaries, one researcher first prepared the summaries by randomly sorting them for each video. Another researcher, unaware of the participant author or interface condition of each summary, graded the participant summaries using the same human-generated summaries used to analyze coverage (Section 5.2). }
Another researcher read interview transcripts to derive themes.

\subsection{Results}
Overall, all participants reported that they would prefer to use \sysname{} to watch short-form videos over the baseline interface. Their average willingness to use \sysname{} in the future was 6.7 $(\sigma=0.67)$ on a scale from 1 (not likely to use in the future) to 7 (very likely to use in the future). Participants expressed that \sysname{} would broaden their access to a wider range of videos and improve the viewing experience: \textit{``this makes me feel like I could view more videos''} (P3), \textit{``it would give me a whole new avenue [...] I would even pay for it''} (P6). Participants also rated \sysname{} on a scale from 1 (not useful) to 7 (very useful) on how useful they found it for both understanding and selecting videos (Figure~\ref{fig:feature-ratings}). For both of these questions, participants rated \sysname{} as significantly more useful compared to the baseline interface: video comprehension ($\mu={6.5}$, $\sigma={0.71}$ vs. $\mu={2.4}$, $\sigma={0.97}$; $Z=2.78$, $p<0.01$) and video selection ($\mu={5.1}$, $\sigma={1.73}$ vs. $\mu={2.3}$, $\sigma={1.42}$; $Z=2.63$, $p<0.01$). Significance was measured with the Wilcoxon Signed Rank test.\\
\textbf{Improved Video Comprehension.}  The accuracy of participant-written summaries significantly improved when using \sysname{} compared to the baseline interface ($\mu = {73\%}$, $\sigma = {26\%}$ vs. $\mu = {20\%}$, $\sigma = {30\%}$; $Z = 4.61$, $p < 0.01$) (Figure~\ref{fig:video-understanding}, left). 
Participant's self-reported video understanding also significantly improved when using \sysname{} compared to the baseline interface ($ \mu = 5.89, \sigma = 1.53$ vs.
$\mu = 2.53, \sigma = 1.93; Z = 4.99, p < 0.01$)  (Figure~\ref{fig:video-understanding}, right).
\revision{When watching videos with the baseline, participants relied heavily on the audio and original video caption for information in the video.
For example, participants watching V6 (a video of a cat with a SpongeBob audio asking a best pet friend to come back) with the baseline interface summarized the video using the topic of the audio (\textit{e.g.}, P4 summarized \textit{``looking for a person or a pet who is lost''}). With ~\sysname{}, however, participants accurately summarized the content of the video despite the audio mismatch (\textit{e.g.}, P9 summarized \textit{``there is a cat and he stole a straw...''}) (Figure 7, V6).}
When using the baseline, participants also tried to use the original video caption for information on the video but reported that \textit{``[captions] are just so variable''} (P3) as \textit{``sometimes [captions are] really useless, other times [they are] good.''} (P10). P3 explained how creators are often more concerned with creating an intriguing hook than a descriptive summary when writing the caption: \textit{``usually they are not as helpful because people are trying to draw attention, and put in hashtags.''} As a result, the caption can also mislead participants.
For example, the caption for video 1 mentioned funny animals even though no animals were present in the video, causing 4 out of the 5 participants who saw this video with the baseline interface to falsely include an animal in their summary. 
With \sysname{}, participants spent more time reading the descriptions and less time reading the caption.\\
\textbf{Using \sysname{}'s Descriptions.}
All of the descriptions provided by \sysname{} (long, short, shot-by-shot and on-screen text) were ranked higher in terms of usefulness when compared to the baseline information (original caption, engagement numbers, audio source, username) (Figure~\ref{fig:feature-ratings}).
Participants reported that each type of description has a unique purpose. 

The short description scored an average of 5.7 ($\sigma = {1.06}$) on a 7-point scale on how useful they found it with 3 participants (P2, P6, P9) giving it a perfect 7 (Figure~\ref{fig:feature-ratings}). 
Participants reported that the short description was helpful for gaining a concise overview of the video to assess whether they were interested in watching the video and/or exploring additional descriptions.
P9 commented that it offered a \textit{``brief glance about what it's gonna be about,''} which she could use to determine if it is worth investigating the video further. P2 reported that she used it as a way to \textit{``measure her interest''} and would use it to \textit{``decide whether [she is] going to look into more detail.''} ~\revision{During the ~\textit{Video Selection Task}, we observed how participants used the descriptions flexibly, only accessing the descriptions when they thought it was worthwhile. P9, for example, would often only read the short description so as to keep the browsing experience low effort. When watching a video of a haircut with audio from a popular song, she read the short description to find out the topic of the video and commented that \textit{``I'm not really interested in hair videos so I'll just skip this.''}}

Participants rated the long descriptions as 6.2 ($\sigma = {0.89}$) on the 7-point scale on how useful they found it (Figure~\ref{fig:feature-ratings}). Participants who rated the long descriptions a 7 (P1, P2, P4, P6, and P8) reported that they liked how the long description \textit{``seemed to cover more of what was done in the video''} than the short description (P8) and provided a good balance between detail and conciseness. 

Participants reported that the on-screen text was particularly useful when videos contained important text information, such as ingredients in a recipe (V3 and V7) or a key joke (V1 and V5). When participants watched a video with important text using the baseline interface, they were unaware of the information they were missing and responded with frustration once we told them what they had missed in the video. P10 found the on-screen text particularly useful and saw it as \textit{``a personal touch''} from the creator of the video.

Participants rated the shot-by-shot descriptions a 6 ($\sigma=1.25$) on the 7-point usefulness scale (Figure~\ref{fig:feature-ratings}).
Participants reported that these descriptions offered them the most detailed and sequential narrative of the video. 
P6 commented that she used shot-by-shot descriptions to \textit{``get the sequence and the whole message''} of the video. Additionally, P4 mentioned that the shot descriptions were a \textit{``broken down''} version of the video and made her realize that \textit{``a lot of stuff was packed into that video that I wouldn’t have known.''} 

Overall, participants found value in the descriptions provided by ~\sysname{} and reported benefits beyond improved video comprehension and selection. P3 shared that ~\sysname{} \textit{``helps with the frustration of saving a video to show someone later''} ~\revision{as the video descriptions often answered questions she had about the video that she might typically save to ask a sighted person to answer later.} P4 commented that even for videos considered more accessible, she could use the descriptions to quickly confirm that she was not missing any information. For example, when watching a video of a woman singing in task 2, she reported that the descriptions were helpful in confirming her assumption that there was a person singing and the audio was not from an outside source.\\
\textbf{Unknown Errors.} While using \sysname{}, participants encountered some errors in the descriptions, often without being aware of them. This happened when descriptions mentioned objects that were not present in the video but resembled objects that were. For example, during task 2, one video featured a kitchen gadget for cutting watermelon, but one of the shot-by-shot descriptions incorrectly identified the watermelon as meat. This led P4 to express surprise, saying \textit{``I never would've guessed there would be meat at the end.''} P1 had a similar experience while watching Video 6 (Table~\ref{tab:user-study-videos}). In this video, the creator addresses the camera directly, but one of the descriptions stated that she is \textit{``interacting with TikTok on her phone.''} As a result, P1 and P2 reported that they thought there was a woman watching a TikTok on her phone rather than speaking to the camera. \\
~\revision{
\textbf{Selecting Descriptions to Read.} Participants appreciated the flexibility of choosing which descriptions to read, allowing them to establish a balance between the effort they are willing to invest and their desire for more information. P9 commented that \textit{``if there’s too much information, its overwhelming''}. They further explain how the short description helped with this issue, saying that \textit{``I felt like I didn't have to go look in more detailed descriptions because [the short description] was so detailed''}. 
Some participants reported that they preferred more detailed descriptions even though they were longer and more time consuming. 
P6 ranked the most detailed description the highest and explained how it \textit{``used details from the videos very well and didn't go into to too much detail''}. Participants would not only choose what description to read, but would also occasionally leave a description if they were not interested in it (\textit{e.g.}, noticed a shot-by-shot description became repetitive). 
We observed that some participants appreciated the process of thoroughly exploring the descriptions and found that reading all the supplemental information, regardless of how accessible the video is, contributed to their understanding of the content. 
Others were less inclined to read the descriptions and used the flexibility of ~\sysname{} to only access descriptions based on their needs (\textit{e.g.}, reading descriptions until their questions were answered).} 
\\
\textbf{Interaction Improvements.} Participants reported potential improvements for \sysname{}, one of which was too much repetition of the same content across the 4 descriptions. Participants also requested: access to the length of the video (P7), separating shot-by-shot descriptions into distinct text components so that they can be read separately by a screen reader (P7 and P10), reformatting the original caption hashtags to be compatible with screen reader (P10), and enabling auto-pause instead of auto-play while scrolling (P8 and P9).
\section{Discussion}
Our formative study revealed that BLV viewers wanted access to visual information in short-form videos (\textbf{D1}, provide visual access). While BLV viewers occasionally use video captions, comments, asked friends or posted on meme pages to gain additional information, existing approaches do not address the accessibility gap. Author-written captions often do not describe visuals, and asking for external help conflicts with the fast pace of short-form videos (\textbf{D5}, maintain fast-paced viewing and browsing). 
\revision{Our hierarchical video summaries supported user study participants in gaining access to on screen visuals, improving self-reported video understanding and video summary scores in our \textit{Video Comprehension Task} (\textbf{D1}). 
Our formative study revealed that audio-visual mismatches led to misconceptions about the visual content (\textbf{D2}, recognizing audio and visual mismatches). 
While participants used \sysname{} to clarify their understanding for audio-visual mismatches (\textbf{D2}), they also used \sysname{} to recognize mismatches between the author-written captions and the visuals.
\sysname{} presents the short descriptions on the video pane (whereas users need to click to the description pane for access to additional descriptions) and thus the short descriptions supported participants in making quick decisions about whether to spend more time on the video or quickly move past it (\textbf{D5}). While participants in the user study \textit{Video Selection Task} occasionally used the short description to decide whether to watch the video (\textbf{D4}, support users in deciding on a video to watch), they also used the short description to decide whether or not to access more visual information about the video.
While the short video descriptions supported efficiency in decision-making and clarification tasks (\textbf{D5}), participants reported that longer descriptions (long, shot-by-shot, and on-screen text) were more useful to fully understanding the visual content (Figure~\ref{fig:feature-ratings}) (\textbf{D1}). 
Based on the findings of our formative study, we enabled screen reader users to have explicit control of the video in our interface and user study participants were able to control videos with ease (\textbf{D3}, screen reader control of video playback and browsing). 
The high coverage (Table~\ref{tab:hallucination-stats}) and low number of errors for descriptions (Figure~\ref{fig:hallucinations}) make them immediately useful such that all users reported wanting to use the descriptions in the future.}
We reflect on our findings to discuss opportunities for future work: 

\subsection{Impact of Generative Model Performance}
Descriptions generated by ~\sysname{} achieved high accuracy, but still contained errors (Figure~\ref{fig:hallucinations}).
While these errors could lead to misconceptions that BLV viewers may not be able to verify, our studies demonstrated that current short-form videos themselves often lead to misconceptions due to mismatched audio and video as well as misleading author-written captions. Short-form videos may represent a relatively low risk avenue for adopting generated descriptions.
\revision{Our summarization-based approach that integrated information across multiple modalities (\textit{e.g.}, transcribed speech and visual descriptions) often removed errors that occurred in earlier stages of the pipeline (\textit{e.g.}, transcription errors). Occasionally, the model would amplify the errors (Figure~\ref{fig:error-analysis}). While participants did not directly express concerns about such errors in the user study, care should still be given to reduce errors in the future as large models can err in the direction of bias~\cite{hendricks2018women}, and misunderstandings of video content have the potential to leave BLV audience members out of the loop (\textit{e.g.}, reacting incorrectly to a video sent by a friend due to a misinterpretation).}
\revision{In the future, we will explore how to reduce such errors by substituting the vision to language model in our pipeline (BLIP-2) with recently released vision to language models (\textit{e.g.}, GPT-V, Google Bard). We will also explore how to mitigate the impact of such errors by providing transparent access to detailed intermediate model outputs, similar to prior work~\cite{huh2023genassist,lee2022imageexplorer}.} For example, future work could enable users to click on a description that seems potentially erroneous to reveal the original 5 BLIP-2 generated image descriptions and their CLIP scores. A high variation of the initial descriptions may prompt skepticism. Future work could alternatively use the CLIP score directly to reveal confidence~\cite{macleod2017understanding}. 

~\revision{
\sysname{} lets users flexibly access descriptions of different levels of detail (\textit{e.g.}, by reading only the short description, or reading all of the descriptions). However, our one-size-fits-all pipeline does not allow customization of the description content.
As a result, descriptions lack content that might be valuable and enjoyable to some users and trivial to others (\textit{e.g.}, detailed description of outfits). 
Future work may explore generating multiple different types of descriptions (\textit{e.g.}, fashion, background descriptions) that users could navigate between or toggle on and off. Alternatively, future work may explore letting users directly customize summarization prompts (\textit{e.g.}, ``leave out any information about color'').

Finally, our pipeline produces description redundancies, both with the original audio of a video and between descriptions~\sysname{}. For example, given that both the short and long descriptions provide a summary of the video, the long description often repeats information in the short description. 
Descriptions also occasionally described the audio of the video unnecessarily (\textit{e.g.}, the description segment ``There is no accompanying audio'', or ``A voice referencing a car game'').
In the future, we could explore filtering out redundancies by further modifying the prompt or comparing the descriptions to the transcript. 
Our user study revealed, however, that some users prefer being given as much information as possible (\textit{e.g.}, detail about the audio can support quickly remembering context of visual details). Thus, future systems should support users to decide whether to allow repetition or not. }
~\revision{\subsection{Extending Support for Short-Form Videos}}
\revision{
Short-form videos showcase a wide degree of variability in length, content theme, and audio type. The variability impacts what descriptions will be useful for what videos. \sysname{} enabled viewers to flexibly scroll on after reading the short description for short videos (\textit{e.g.}, a dog jumps into a pool) and flexibly gain additional access for informative videos (\textit{e.g.}, a recipe). While on-screen text was valuable for jokes and information, it was not useful when directly transcribing the audio. 
A benefit to our approach of large language model generated summaries, is that the summaries often suppressed redundant information such as a matching transcript and on-screen text making the descriptions work across a range of videos. Similarly, image-based vision to language models (\textit{e.g.}, BLIP-2, GPT-V) perform well across a range of visuals. 
Still, ~\sysname{} lack support for specific types of videos such as videos with complex actions (\textit{e.g.}, dance, trick shot challenges) or reaction videos (\textit{e.g.}, a side-by-side view of a video and a person reacting to the video, or a 2x2 view of four musicians playing together) due to limitations of our pipeline and limits of descriptions. 

\textbf{Improving our Generative AI Pipeline:} We chose to the vision to language model BLIP-2 in our pipeline due to its high performance across diverse visuals. However, BLIP-2 takes single images as input and thus lacks information to recognize complex movements across time. In the future, we expect the performance of video to language models to improve, and we will also explore specialized pipelines such as accurately reconstructing 3D meshes of humans during dances such that the motions are robust to occlusion~\cite{goel2023humans} then describing the 3D meshes~\cite{delmas2022posescript,yamada2018paired}. Vision to language models also struggle to recognize and accurately describe structured videos such as reaction videos. One approach may be to classify and segment reaction videos, create hierarchical descriptions for each individual video, then create a short description that summarizes all of the videos as a whole to provide an overview. 

\textbf{Alternative Description Modalities:} Generating audio descriptions for complex actions in a limited time is a well-known challenge. Dance, for example, contains rapid and complex movements as well as facial expressions to convey emotional tone. Describing such movements in real-time without overwhelming the listener is difficult and can often take BLV audiences out of the experience of enjoying the performance~\cite{danceinaccessible}. Describing subjective themes such as emotion also violates audio description guidelines ~\cite{acbguidelines} which state that descriptions should be made \textit{``without interpretation or personal comment.''} 
Some suggestions have been made to improve the accessibility of dance through haptic tours that allow BLV audiences to physically touch the set and costumes before the performance ~\cite{curtisdanceaccessible}.
Haptics could also be used to augment descriptions with rich spatial information~\cite{viswanathan2010haptics}.
When the purpose of the dance video is a tutorial rather than a performance (\textit{e.g.}, a short video on how to do a popular dance move), future work may explore giving detailed auditory-only instructions and feedback as Rector et al. provided for yoga~\cite{ractor2013yoga}. Sonification has also been used to improve accessibility of spatial navigation~\cite{onofrei2023sonification} and data visualization~\cite{holloway2022infographics} by representing certain attributes (\textit{e.g.}, distance and data trends) as specific audio queues. However, haptics and sonification are time-based modalities (i.e. they play alongside the video as it plays) and may be challenging to integrate with short-form videos that contain rapid visual changes.  
}\\

\subsection{Platform Recommendations}
The increased popularity of short-form videos supports the need for major platforms (i.e. TikTok, Instagram Reels, and YouTube Shorts) to leverage their access to advanced technology to improve accessibility. In the formative study, participants shared their difficulty in accessing both short-form video content and the platforms. 
We explored \sysname{} as an approach for making short-form videos accessible. Our work reveals the following design implications:

\noindent \textbf{AI Descriptions to Fit Information Needs.} While participants found short descriptions beneficial for scrolling (\textit{e.g.}, analogous to viewing the first few frames for sighted viewers), additional information is required for videos of interest. Hierarchical visual description summaries support flexible information access. 

\revision{\noindent \textbf{Prioritizing Useful Details.} In our user study, participants rated the username as the least useful piece of information for understanding a short-form video, yet this is typically the first piece of information presented to them on short-form video platforms.
For efficiency, the screen reader order should reflect what will best support BLV users in understanding the content (\textit{e.g.}, the caption for current interfaces).}

\noindent \textbf{Screen Reader Access.} Platforms also have the ability to make small adjustments that can significantly improve BLV users' experience watching short-form videos. For example, platforms may consider making on-screen text added using their platform screen reader accessible. 
Video playback and browsing should be improved by providing easy access to pause to avoid overlapping short videos or enabling ``auto-pause'' as an alternative to ``auto-play.''

\noindent \textbf{Short-Form Video ``Alt-Text'' Field. } BLV viewers reported that author-written captions are often used to complement sighted-viewers viewing experience and extend view times rather than describe visual details. \sysname{}'s short descriptions were helpful to clear up misunderstandings. While some content creators use the caption field to add a visual description~\cite{tiktok-descriptions}, this is not the norm. An ``alt-text'' field for authors to write visual descriptions (similar to Instagram and Twitter) could support authors in making their content accessible.

\revision{\noindent \textbf{Short-Form Video Accessibility Guidelines} Web Content Accessibility Guidelines (WCAG 2.0) recommend that to make videos accessible, authors describe the content of a video or provide synchronous audio descriptions to narrate visual content~\cite{wcagad}. Accessibility guidelines, however, have not been established specifically for short-form videos. Our formative study established unique challenges that make short-form videos different from traditional videos, such that new guidelines would be beneficial.}

\noindent \textbf{Awareness. } Currently, short-form video content creators are not encouraged to consider a BLV audience while creating their content. 
Creators describing the visual content during production or adding descriptive text could improve short-form video accessibility.

~\revision{\subsection{Beyond Short-Form Videos}
Our work demonstrates the potential to use a mix of visual information extraction via vision language models and summarization via large language models to create useful visual descriptions at multiple levels of detail. Prior work explored a similar approach for extracting and summarizing image details~\cite{huh2023genassist}, but our work demonstrates the potential of this approach for temporal media. In the future, we can explore extending ShortScribe to long-form videos, livestream recordings, or 360-degree videos. The hierarchical summaries may also support other tasks. For instance, prior work revealed that BLV people want adaptive length descriptions for digital comics~\cite{huh2022cocomix}, shopping images~\cite{stangl2018browsewithme}, and urban scenes~\cite{hoogsteen2022beyond}. We hope our work catalyzes future research exploring accessible descriptions that fit diverse user contexts and preferences. 
}

\section{Conclusion}
In this paper, we presented ~\sysname{}, a system created to make short-form videos accessible to blind and low vision (BLV) viewers. Supported by findings from the formative study, ~\sysname{} provides on-demand descriptions covering various levels of detail generated by leveraging vision language and large language models. We evaluated the effectiveness of ~\sysname{} through a technical evaluation of description accuracy and coverage, and a user study with 10 BLV participants. \revision{In the user study, participants reported higher comprehension and provided more accurate summaries of video content. Participants flexibly navigated between different descriptions, and found all descriptions to be useful for different purposes. All participants stated they wanted to use \sysname{} in the future.} We aim to motivate future work to support people with disabilities in engaging with and creating this new growing form of social media. 



\bibliographystyle{ACM-Reference-Format}
\bibliography{sample-base}

\appendix

\section{Appendix - GPT-4 Prompts}
This section of appendix includes the prompts given to GPT-4 which are referenced in the System section of the paper.\\

\subsection{GPT-4 Long Description Summarization Prompt} \label{long_desc_prompt}

Your task is to generate a summary paragraph for an entire short-form video based on data extracted from the video. Your summary must be a holistic description of the full video.\\ \\
The text in quotations defines the format of the data that I will provide you. The video data comprises of data extracted from all shots of the video.\\ \\
The data is formatted in the structure defined in the quotations: \\
"SHOT NUMBER\\
Duration: the number of seconds that the shot lasts\\
Text on screen: Any text that appears in the shot\\
Shot audio transcript: Any speech that is in the shot\\
Shot description: A short visual description of what is happening in the shot"

\subsection{GPT-4 Shot-by-Shot Description Summarization Prompt} \label{shot_by_shot_prompt}

Your task is to generate a summary for each shot of a short-form video based on data extracted from the video.\\ \\
The text in quotations defines the format of the data that I will provide you. The video data comprises of data extracted from all shots of the video.\\ \\
The data is formatted in the structure defined in the quotations:\\
"SHOT NUMBER\\
Duration: the number of seconds that the shot lasts\\
Text on screen: Any text that appears in the shot\\
Shot audio transcript: Any speech that is in the shot\\
Shot description: A short visual description of what is happening in the shot"\\\\
All of the summaries you create must satisfy the following constraints:\\
1. If the field for text on screen is empty, do not include references to text on screen in the summary.\newline
2. If the field for shot audio transcript is empty, do not include references to shot audio transcript in the summary.\newline
3. If the field for shot description is empty, do not include references to the shot description in the summary.\newline
4. If the field for shot description is empty, do not include references to shot description in the summary.\newline
5. Do not include references to Tiktok logos or Tiktok usernames in the summary.\\\\
Provide the summaries in a newline-separated format. There must be exactly one summary for every shot.\\\\
You must strictly follow the format inside the quotations.\\
"Your first summary\newline
 Your second summary\newline
 Your third summary\newline
 More of your summaries...\newline
 Your last summary" \\

\subsection{GPT-4 50-word Description Prompt} \label{50wordprompt}
``Condense the summary below such that the response adheres to a 50 word limit.''




\section{Appendix - Participant and Video Data}
The following appendix section contains participant and video data from the formative study and user evaluation. 

\begin{table*}[h]
\centering
\resizebox{5in}{!}{\begin{tabular}{>{\raggedright\arraybackslash}m{0.3cm} >{\raggedright\arraybackslash}m{0.3cm} >{\raggedright\arraybackslash}m{0.8cm} >{\raggedright\arraybackslash}m{1.8cm} >
{\raggedright\arraybackslash}m{1cm} >{\raggedright\arraybackslash}m{1.8cm} >{\raggedright\arraybackslash}m{2.5cm} >{\raggedright\arraybackslash}m{1.5cm} >{\raggedright\arraybackslash}m{2 cm} >{\raggedright\arraybackslash}m{2.3cm}} 
\toprule
ID & Age & Gender & Vision Impairment & Onset Age & Assistive Technology & Short-Form Video Platform(s) & Amount of Experience & Frequency 
& Preference for the type of videos\\ \midrule
P1 & 29 & Female & L: Totally Blind;
R: Legally Blind
& 1 & VoiceOver, Zoom text & YouTube, Instagram & 1 year & 4 times per day & Fashion, Informative \\ 
P2 & 33 & Male & Totally Blind & 11 &VoiceOver, Jaws &TikTok, YouTube, Facebook & 2 or 3 years & 10 hrs per day
& Informative \\
P3 & 32 & Male & Registered Blind & Teen & VoiceOver, Jaws, NVDA & TikTok, Instagram, YouTube, Facebook & 3 years & Daily & Informative \\
P4 & 29 & Female & Legally Blind & 12 & VoiceOver, Jaws & YouTube, Facebook
& < 1 year
& 1-3 times per week & Comedy \\
P5 & 43 & Female & Totally Blind & 1 & VoiceOver & TikTok & 2 years & 15-20 mins per day & Shorter videos\\
P6 & 57 & Male & Totally Blind & 1.5 & NVDA & TikTok & 4 years & 1-2 times per week & Informative, Live Music\\
P7 & 20 & Male & Legally Blind & Birth & VoiceOver & TikTok, YouTube, Facebook, Instagram & 4 years & 
1 time per week & Informative\\
\bottomrule
\end{tabular}}
\caption{Background of Participants in Formative Study. L- left eye, R - right eye}
\label{tab:pbackground-formative}
\end{table*}

\begin{table*}[h]
\resizebox{5in}{!}{\begin{tabular}{>{\raggedright\arraybackslash}m{1cm} >{\raggedright\arraybackslash}m{1cm} >{\raggedright\arraybackslash}m{7cm} >{\raggedright\arraybackslash}m{7cm} }
\toprule
VID & Length  & Visual Content & Audio Content\\ \midrule
V1 & 23s & Woman telling her small dog not to bark at larger dogs when at the park & Woman speaking to her dog \\
V2 & 1m 17s & Woman singing a song to the camera & Woman singing song\\
V3 & 5s & Cat yawning on a bed while laying down & Meme-style audio about how "today drained me." \\
V4 & 36s & Woman shows the steps to making a salad often made by a popular actress & Popular upbeat song \\
V5 & 5s & Dog jumping in pool even though its owner tries to keep him from doing so & Background noise of dog jumping in pool \\
V6 & 40s & Woman giving a detailed step-by-step recipe on how to make a pasta dish & Woman explaining how to do each step\\
V7 & 5s & Dog showing off its stuffed toys while laying on the floor & Meme style audio referencing "little babies" \\
V8 & 59s & A couple taking a quiz that had been trending & The wife asking the husband questions and him answering\\
\bottomrule
\end{tabular}}
\caption{Content of 8 Pre-selected Videos from Formative Study}
\label{tab:pre-selected videos}
\end{table*}

\begin{table*}[h]
\centering
\resizebox{5in}{!}{\begin{tabular}{l l l l l l}
\toprule
PID & Gender & Age & Visual Impairment & Onset & Prefered Short-From Video Platform (If any)\\ \midrule
P1 &Female &45 &Legally blind &Congenital &YouTube shorts \\
P2 &Female &43 &Totally blind &Congenital &TikTok \\
P3 &Female &29 & Light perception &Acquired &Instagram Reels, YouTube Shorts \\
P4 &Female &29 &Light perception &Congenital &YouTube Shorts \\
P5 &Female &33 &Totally blind &Congenital &YouTube, Instagram, Facebook, Tiktok \\
P6 &Female &63 &Totally blind &Acquired &YouTube Shorts \\
P7 &Male &32 &Light perception &Acquired &YouTube Shorts \\
P8 &Female &68 &Light perception &Congenital &None \\
P9 &Female &29 &Legally blind &Acquired &Youtube Shorts, Facebook Shorts \\
P10 &Male &27 &Totally blind &Acquired &Youtube Shorts \\
\bottomrule
\end{tabular}}
\caption{Background of Participants in User Study.}
\label{tab:pbackground-userstudy}
\end{table*}

\begin{table*}[h]
\resizebox{5in}{!}{\begin{tabular}{>{\raggedright\arraybackslash}m{0.5cm} >
{\raggedright\arraybackslash}m{0.7cm} >{\raggedright\arraybackslash}m{1cm} >{\raggedright\arraybackslash}m{6.5cm} >{\raggedright\arraybackslash}m{6.5cm} } 
\toprule
VID &Group &Length &Visual Content &Audio Content \\\midrule
V1 &1 &18s &A dad on a conference call falls in a pool while and continues the call as if nothing happened &Original audio from video \\
V2 &1 &10s &Zoom in of a streat sign that reads "Drury Ln" &Man exlaims "Oh my gosh, you guys...The Muffin Man" \\
V3 &1 &23s &A man preparing vegan mozarella &The video is introduced followed by a song overtop background noise of preparing the recipe \\
V4 &1 &54s &A man gives a list of 5 songs he believes should never be played on the accoustic guitar while playing them on his guitar &The video is introduced followed by the man playing potions all 5 of the songs on the guitar \\
V5 &2 &13s &A dog that escaped home rings the front door bell with his nose &Original audio from video \\
V6 &2 &8s &A cat is shown stealing a straw from a cup with the sentiment that the creator loves the cat regardless &Meme audio originally from SpongeBob about pets being your best friend \\
V7 &2 &33s &A tutorial-style video on how to prepare tato corn chow &The audio is a fantasy style rendition of a popular song \\
V8 &2 &25s &A woman gives a list of 5 basic italian phrases that everyone should know &The video is introduced followed by the woman pronouncing all 5 of the phrases \\
\bottomrule
\end{tabular}}
\caption{Content of 8 pre-selected videos from User Study.}
\label{tab:user-study-videos}
\end{table*}

\end{document}